\documentclass[superscriptaddress,onecolumn]{revtex4-2}
\usepackage{graphicx,hyperref}
\usepackage{mathrsfs,bbm,amsmath,amssymb,times}
\usepackage{mathtools}
\DeclarePairedDelimiter\floor{\lfloor}{\rfloor} 
\usepackage{multirow}
\def\Dk{{\rm dc}}
\begin{document}

\title{An Enhanced Photonic Quantum Finite Automaton}

\author{Alessandro Candeloro}
\affiliation{Dipartimento di Fisica ``Aldo Pontremoli'', Universit\`a degli  Studi di Milano, I-20133 Milano, Italy}

\author{Carlo Mereghetti}
\affiliation{Dipartimento di Fisica ``Aldo Pontremoli'', Universit\`a degli  Studi di Milano, I-20133 Milano, Italy}

\author{Beatrice Palano}
\affiliation{Dipartimento di Informatica ``Giovanni Degli Antoni'', Università degli Studi di Milano, via Celoria 18, I-20133 Milano, Italy}

\author{Simone Cialdi}
\affiliation{Dipartimento di Fisica ``Aldo Pontremoli'', Universit\`a degli  Studi di Milano, I-20133 Milano, Italy}
\affiliation{INFN, Sezione di Milano, I-20133 Milano, Italy}

\author{Matteo G.~A.~Paris}
\affiliation{Dipartimento di Fisica ``Aldo Pontremoli'', Universit\`a degli  Studi di Milano, I-20133 Milano, Italy}
\affiliation{INFN, Sezione di Milano, I-20133 Milano, Italy}

\author{Stefano Olivares}
\email{Electronic address: stefano.olivares@fisica.unimi.it}
\affiliation{Dipartimento di Fisica ``Aldo Pontremoli'', Universit\`a degli  Studi di Milano, I-20133 Milano, Italy}
\affiliation{INFN, Sezione di Milano, I-20133 Milano, Italy}

\begin{abstract}
In a recent paper we have described an  optical implementation of a \emph{measure-once one-way quantum finite automaton} recognizing a well-known family of {unary} periodic languages, accepting words not in the language with a given error probability. To process input words, the automaton exploits the degree of polarization of single photons and, to reduce the acceptance error probability, a technique of confidence amplification using the photon counts is implemented. In this paper, we show that the performance of this automaton 
may be further improved by using strategies that suitably consider {\em both} the orthogonal output polarizations of the photon. In our analysis, we also take into account how 
detector dark counts may affect the performance of the automaton.
\end{abstract}


\maketitle

\section{Introduction}
In the recent years, quantum computers have eventually leaped out of the laboratories \cite{Castelvecchi17} and become accessible to a still growing community interested in investigating their actual potentialities. Nevertheless, a full-featured quantum computer is still far from being built. However, it is reasonable to think of classical computers exploiting some quantum components. In this framework, quantum finite automata \cite{ambainis2018automata,bhatia2019}---theoretical models for quantum machines with finite memory---may play a key role, since they model small-size quantum computational devices that can be embedded in classical ones. Among possible models, the so-called measure-once one-way quantum finite automaton \cite{BERTONI2001174,Brodsky02} is the simplest and it has been shown to be the most promising for a physical realization \cite{PhysRevResearch.2.013089}. In fact, restricted models of computation, such as quantum versions of finite automata, have been theoretically studied \cite{BERTONI2005394,MEREGHETTI2007177,AMBAINIS20091916} and, very recently, experimentally investigated \cite{PhysRevResearch.2.013089,Birkan2021}.

In~\cite{PhysRevResearch.2.013089}, a measure-once one-way quantum finite automaton recognizing a well-known family of unary periodic languages \cite{BERTONI2001174}, namely languages $L_m$, has been implemented using quantum optical technology \cite{CIaldi17,Olivares21}. In our implementation, a given input word is accepted by the automaton, with a given error probability, whenever a single photon arrives at the output of the device with a specific polarization. In particular, the experimental realization, based on the manipulation of single-photon polarization and photodetection, have demonstrated the possibility of building small quantum computational component to be embedded in more sophisticated and precise quantum finite automata or also in other computational systems and approaches \cite{de_Falco_2013,Tamascelli_2014,Rossi_2017}. Albeit the photonic automaton realized in \cite{PhysRevResearch.2.013089} is fed with single photons, it works in a regime where polarized laser pulses (coherent states) are enough, up to detecting the intensity of the output signals instead of counting the number of photons successfully passing through the device with a given polarization (see \cite{PhysRevResearch.2.013089} for details).

In this paper, we propose an enhanced version of our photonic automaton mentioned above, where, to further reduce the acceptance error probability, we consider not only the photons with the ``correct'' polarization, but also the other ones. To achieve this goal, the use of single-photon techniques turns out to be crucial, such as the detection of coincidence count to reduce the dark-count rate of the photodetectors \cite{bachor}. Analytical and numerical results, supported by simulated experiments, show that the enhanced version allows to reduce the error probability by orders of magnitude compared  to the previous version, or, analogously, to drastically reduce the mean number of photons needed to achieve the same performance.

The paper is structured as follows. Since our work requires some previous knowledge from Theoretical Computer Science about formal languages and finite automata, section~\ref{s:motivations} is devoted to introduce the reader to these topics, providing the relevant motivations. In section~\ref{s:1qfa} we briefly review basics of formal language theory and the definition of a measure-once one-way quantum finite automaton. Section~\ref{s:implem} describes the implementation of the measure-once one-way quantum finite automaton based on the polarization of single photons, linear optical elements and photodetectors. In section~\ref{s:enhance} we explain how to improve the confidence of the obtained measure-once one-way quantum finite automaton by processing the number of counts at the detectors. We also introduce new strategies that reduce the error probability, namely, the probability that a ``wrong'' word is accepted by the automaton or a ``correct'' word is rejected. The numerical results and the simulated experiments are reported in Section~\ref{s:numerical}. We close the paper with some concluding remarks in section~\ref{s:concl}.

\section{Formal languages, finite automata and quantum computing}\label{s:motivations}
In this section we would like to expand on motivations that have been driving our research covered by the present contribution and the previous one in \cite{PhysRevResearch.2.013089}. The aim of our work, that bridges between Theoretical Computer Science and Experimental Quantum Optics, has been and is to show that a quantum computing device with finite memory is physically realizable by means of photonics, using a very limited amount of ``quantum hardware''.
To the best of our knowledge, our physical implementation, described here and in \cite{PhysRevResearch.2.013089}, of a quantum finite automaton for language acceptation is the first proposed in the literature.
Thus, we have shown how the quantum behaviour of microscopic systems can actually represent a computational resource, as theoretically established within the discipline of Quantum Computing.
From this perspective, the simple language $L_m$, introduced in the next section and for which we build our photonic quantum finite automaton, is not really the point here. Instead, the point is the concrete creation of a programmable fully quantum computer with finite memory.

With this being said, we would also like to quickly comment on the language~$L_m$ from a Theoretical Computer Science viewpoint. Notwithstanding its simplicity, the language $L_m$ plays a crucial role in Descriptional Complexity Theory (see, e.g., \cite{BGMP14,BMPP11,CMMP12,JMMP13,KMMP19}), the area of Formal Language Theory in which the {\em size} of computational models is investigated. In particular, a well consolidated trend in Descriptional Complexity is devoted to study the size of several types of \emph{finite  automata}. The reader is referred to, e.g., \cite{Hopcroft79} for extensive presentations of automata theory.
Very roughly speaking, the hardware of a (one-way) finite automaton $A$ features
a read-only input tape consisting of a sequence of cells, each one being able to store an input symbol. The tape is scanned by an input head always moving one position right at each step. At each time during the computation of $A$, a finite state control is in a state from a \emph{finite} set $Q$. Some of the states in $Q$ are designated as accepting states, while a state $q_0 \in Q$ is a designated initial state. The computation of~$A$ on a word (i.e, a finite sequence of symbols) $\omega$ from a given input alphabet begins by having (i) $\omega$ stored symbol by symbol, left to right, in the cells of the input tape, (ii) the input head scanning the leftmost tape cell, and (iii) the finite state control being in the state $q_0$. In a move, $A$ reads the symbol below the input head and, depending on such a symbol and the state of the finite state control, it switches to the next state according to a fixed transition function and moves the input head one position forward. We say that $A$ accepts $\omega$ whenever it enters an accepting state after scanning the rightmost symbol of~$\omega$; otherwise, $A$ rejects $\omega$. The language accepted by~$A$ consists of all the input words accepted by $A$.

The one described so far is the original model of a finite automaton, called {\em deterministic}. Several variants of such an original model have been introduced and studied in the literature, sharing the same hardware but different dynamics. So, we have \emph{nondeterministic}, \emph{probabilistic} and, recently, \emph{quantum} finite automata (see, e.g., \cite{BMP10,BMP17,BPa09}). Also, \emph{two-way} automata are studied, where
the input head can move back and forth on the input tape.

Finite automata represent a formidable theoretical model used in the design and analysis of several devices such as the control units for vending machines, elevators, traffic lights, combination locks, etc. Particularly important is the use of finite automata in very large scale integration (VLSI) design, namely, in the project of sequential networks which are the building blocks of modern computers and digital systems. Very roughly speaking, a sequential network is a boolean circuit equipped with memory. Engineering a sequential network typically requires modelling its behaviour with a finite automaton whose number of states directly influences the amount of hardware (i.e., the number of logic gates) employed in the electronic realization of the sequential network. From this point of view, having fewer states in the modelling finite automaton directly results in employing smaller hardware which, in turn, means having less energy absorption and fewer cooling problems. 
These ``physical'' considerations, of paramount importance given the current level of digital device miniaturization, have led to define the {\em size} of a finite automaton as the {\em number of its states}.
In particular, reducing or increasing the number of states is studied, when using different computational paradigms (e.g., deterministic, nondeterministic, probabilistic, quantum, one-way, two-way) on a finite automaton to perform a given task. Here is where our simple language $L_m$ comes into play. In fact, this language is universally used as a benchmark to emphasize the succinctness of several types of automata. 
Several results in the literature shows that accepting $L_m$ on classical models of finite state automata is particularly size-consuming (i.e, it requires a great number of states), while only two basis states are enough on quantum finite automata, as we will see in the next section.

Modular design frameworks have been theoretically proposed \cite{BERTONI2005394,MEREGHETTI2007177,AMBAINIS20091916}, where more reliable and sophisticated quantum automata can be built by suitably composing (see, e.g., \cite{BGMP13}) easy-to-obtain variants of the quantum automaton for $L_m$. Hence, our work provides crucial and concrete quantum components for such frameworks, and suggest investigating a physical implementation of some automata composition laws. More generally, the Krohn-Rhodes decomposition theorem \cite{Ginzburg:atoa:1968} states that \emph{any} classical finite automaton can be simulated by composing very ``simple'' finite automata: one of these simple automata is exactly the one for~$L_m$. From this perspective, our photonic quantum automaton could be hardwired into ``hybrid'' architectures joining classical and quantum components to build very succinct finite state devices operating in environments where dimension and energy absorption are particularly critical issues (e.g., drone or robot-based systems \cite{FMP18}).

\section{Measure-Once One-Way Quantum Finite Automaton}\label{s:1qfa}
Here, we briefly overview main concepts on automata and formal language theory. We refer the interested reader to any of the standard books on these subjects (see, e.g., \cite{Hopcroft79}), as well as to our contribution \cite{PhysRevResearch.2.013089}.

An alphabet is any finite set $\Sigma$ of elements called symbols. A word on $\Sigma$ is any sequence 
$\sigma_1\sigma_2\cdots\sigma_n$ with $\sigma_i\in\Sigma$.
The set of all words on~$\Sigma$  is denoted~by~$\Sigma^*$.
A language~$L$ on $\Sigma$ is any subset of~$\Sigma^*$, i.e., $L\subseteq\Sigma^*$.
If $|\Sigma|=1$ we say that $\Sigma$ is a \emph{unary} alphabet, and languages on unary alphabets are called unary languages. In case of unary alphabets, we customarily let $\Sigma=\{a\}$ so that a {unary language} is any set~$L\subseteq a^*$. We let~$a^k$ be the unary word obtained by concatenating $k$ times the symbol $a$.

In what follows, we will be interested in the unary language $L_m$ defined~as
\begin{equation}
\label{eq:unarylanguage}
L_{m} = \{a^{k}\mid k \in \mathbb{N}\,\mbox{ and }\,k(\mbox{mod}~m) = 0 \}.
\end{equation}
This language is rather famous in the realm of automata theory, since it has proven particularly ``size-consuming''
to be accepted by several models of classical automata, since the number of needed states increases with $m$ \cite{PhysRevResearch.2.013089}. The reader may find a deep investigation on this fact in the literature \cite{Hopcroft79,Paz71,mere00,mere01}. On the other hand, as presented in \cite{PhysRevResearch.2.013089}, very succinct measure-once one-way quantum finite automata (1qfa's, from now on) may be designed and physically realized for $L_m$. Let us now sketch the main ingredients for a 1qfa accepting $L_m$.

If we consider the two orthogonal states $| H \rangle = (1,0)$ and $| V \rangle = (0,1)$, the 1qfa is defined as (here we use the formalism based on the Dirac's notation; the analysis based on a more general formalism can be found in~\cite{PhysRevResearch.2.013089})
\begin{equation}
\mathcal{A}_{1} = \left\{| H \rangle , U_{m}, P^{H} \right\}
\end{equation}
where $| H \rangle$ represents the initial state, the unitary operation applied by the automaton upon processing any input symbol $a$ is defined as
\begin{align}
U_{m} &= \exp(-i \theta_m \sigma_y)\\[1ex]
&=\begin{pmatrix}
\cos\theta_m & \sin\theta_m \\[1ex]
-\sin\theta_m & \cos\theta_m
\end{pmatrix}.
\end{align}
with $\theta_m = \pi/m$ and $\sigma_y$ the Pauli matrix, while $P^{H} = | H \rangle\langle H |$ is the projector onto the mono-dimensional {\em accepting} subspace spanned by $| H \rangle$. The probability $p_{\mathcal{A}_{1}}(a^{k}) $ that the 1qfa $\mathcal{A}_1$ accepts the word $a^{k}$ writes as 
\begin{align}
p_{\mathcal{A}_{1}}(a^{k}) &=  p^{H}(a^{k})  \equiv | \langle H | U_{m}^k | H \rangle  |^2 \\[1ex]
&= \cos^{2}\left(k \theta_m\right) \to
\begin{cases}
= 1 & k (\mbox{mod}~m)= 0 \\
\leq \cos^{2}\theta_m & \textup{otherwise.}
\end{cases}
\label{eq:pA1}
\end{align} 
Therefore, the 1qfa $\mathcal{A}_{1}$ perfectly recognizes the word $a^{k} \in L_{m}$, since we can set a {\it cut point}~$\lambda$ and an {\it isolation} $\rho$ to the following values (see~\cite{PhysRevResearch.2.013089} for details on accepting languages with isolated cut point)
\begin{equation}
\lambda = \frac{1+\cos^{2}\theta_m}{2} \quad \mbox{and} \quad \rho = \frac{1-\cos^{2}\theta_m}{2}\,.
\end{equation}
However, $\mathcal{A}_{1}$ may also recognize an input word not in $L_m$ with a non-null probability. In the following, we let $a^{k_1}$ with \ $k_1 (\mbox{mod}~m) = 1$ any of the word with the highest probability of erroneously being accepted, i.e. $\cos^{2}\theta_m$, which tends to 1 as $m$ gets large. This can be seen also by the fact that $\rho \to 0$ as $m$ increases.
\par
As matter of fact, we can also introduce the following 1qft, where we still consider the initial state $| H \rangle$, but, at the output, we focus on the final projection involving the state $| V \rangle$, namely
\begin{equation}
	\mathcal{A}_{2} = \left\{| H \rangle, U_{m}, \mathbbm{I} - P^{V}\right\},
\end{equation}
where $P^{V} = | V \rangle\langle V |$. Indeed, $\mathcal{A}_{2}$ is formally equivalent to ${\mathcal A}_1$, since $\mathbbm{I} - P^{V} \equiv P^{H}$. In fact, the probability of accepting a word is now given by
\begin{equation}
	p_{\mathcal{A}_{2}}(a^{k}) = 1-p^{V}(a^{k}) \to 
\begin{cases}
= 1 & k (\mbox{mod}~m) = 0 \\
\leq \cos^{2}\theta_m& \textup{otherwise}
\end{cases}
\label{eq:pA2}
\end{equation}
that is the same as in Eq.~\eqref{eq:pA1}, as one may expect. Nevertheless, we show in the next section that the two are not equivalent in a photonic implementation for reasons that will be clear soon.

\section{Photonic implementation of the 1qfa}\label{s:implem}
\begin{figure}
\includegraphics[width=0.5\textwidth]{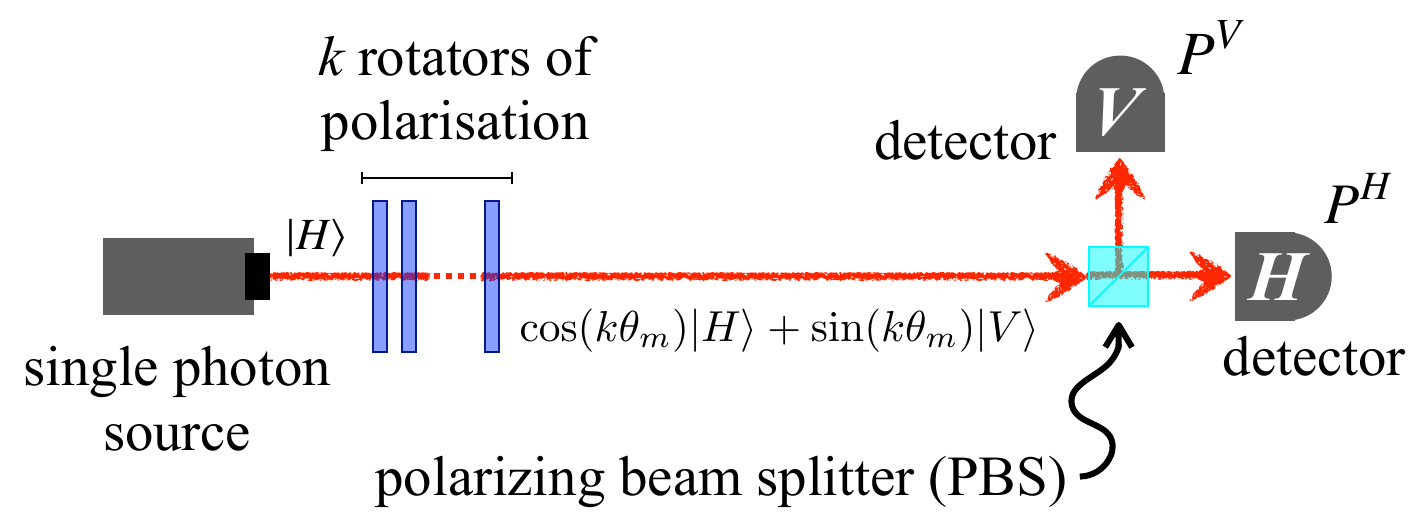}
\caption{Scheme of the photonic implementation of the 1qfa highlighting the main involved optical elements. See the text for details.}
\label{fig:scheme}
\end{figure}
A photonic implementation of the 1qfa described in the previous section was proposed and demonstrated in \cite{PhysRevResearch.2.013089}. Figure~\ref{fig:scheme} depicts the main elements of the enhanced version of the automata we will describe in the following.

The state of the automaton is encoded in the polarization of single photons, and the Hilbert space is $\mathcal{H}=\textup{span}\{\vert H \rangle,\vert V \rangle\}$. A single photon source generates a horizontal-polarized state $\vert H\rangle$, which is sent to $k$ rotators of polarization, $a^k$ being the input word to be processed. Each rotator corresponds to a unitary rotation of an amount $\theta_{m}$, which is thus language-dependent. After the rotators, the single photon state reads
\begin{equation}
\vert k\theta_{m} \rangle = \cos(k\theta_{m}) \vert H \rangle + \sin(k\theta_{m}) \vert V \rangle 
\end{equation}
and it is sent to a polarizing beam splitter (PBS in Fig.~\ref{fig:scheme}) that reflects the vertical polarization component and transimits the horizontal one. Finally, two photo-detectors placed after the PBS realize the projective measure of $P^{H}$ and $P^{V}$. As the reader can see, the scheme is almost the same of that proposed in \cite{PhysRevResearch.2.013089}, but here we will implement a new inference strategy exploiting the outcomes from both the detectors.

\par
As we observed in the previous section, the automata $\mathcal{A}_{1}$ and $\mathcal{A}_{2}$ accept with certainty a word $a^{k}$ that belongs to $L_{m}$. However, there is a high probability that an incorrect word, such as $a^{k_1}$ with $k_1 \mod m \ne 0$ can be accepted, as we can see from Eqs.~\eqref{eq:pA1} and \eqref{eq:pA2}. Hence, strategy based on a single-photon shot may not be the optimal way to recognize an arbitrary word $a^{k}$.

\section{Confidence amplification: an enhanced strategy}\label{s:enhance}
To reduce the probability of error, we can adopt a technique of confidence amplification as also proposed in  \cite{PhysRevResearch.2.013089}, namely, we sent a mean number of photons $\langle N_{\rm c}\rangle$ and we count the number of click $N_{\rm c}^{x}(k)$ at the photodetector $x=H,V$, see Fig.~\ref{fig:scheme}. Therefore, the observed detection frequency at detector $x=H,V$ for an input word $a^{k}$ will be
\begin{equation}
	f_{k}^{x} = \frac{N_{\rm c}^{x}(k)}{\langle N_{\rm c}\rangle} \xrightarrow{\langle N_{\rm c}\rangle\gg 1} p^{x}_{\mathcal{A}_{i=1,2}}(a^{k})\,.
\end{equation}
Thereafter, we turn our problem into that of discriminating among the corresponding detection frequencies and, in particular, we can focus on those related to $k=0$ (or, equivalently, $k \mod m =0$) and $k=1$ (or, more in general, $k \mod m =1$), since if $k >1$ one has $f_{k}^{H} < f_{1}^{H}$ ($f_{k}^{V}> f_{1}^{V}$). To implement this strategy, we set a threshold frequency as 
\begin{equation}
f^{x}_{\rm th} = \frac{f^{x}_{0}+f^{x}_{1}}{2} =
\begin{cases}
	 \displaystyle \frac{1+f^{H}_{1}}{2} & x= H\\[2ex]
	 \displaystyle \frac{f^{V}_{1}}{2} & x= V
\end{cases},
\end{equation}
where $f^{H}_{1}$ ($f^{V}_{1}$) is the highest (lowest) frequency of erroneously accepted words $a^{k_1}$, while $f_{0}^{H}$ ($f_{0}^{V}$) is the frequency corresponding to the correct word. In this formula we have distinguished the two different strategies: for the $H$ detector, $f_{0}^{H}=1$, since the corresponding photon will always be detected; instead, for the $V$ detector, $f_{0}^{V}=0$, since no photon is detected when the word belongs to $L_{m}$. Hence, the strategy is to accept the word if $f_{k}^{H}> f_{\rm th}^{H}$ ($f_{k}^{V}< f_{\rm th}^{V}$) and reject it if $f_{k}^{H}< f_{\rm th}^{H}$ ($f_{k}^{V}> f_{\rm th}^{V}$). From now on, we will refer to these strategies as ``H strategy'' and ``V strategy'', respectively.

\begin{figure}
\centering
\includegraphics[width=0.7\textwidth]{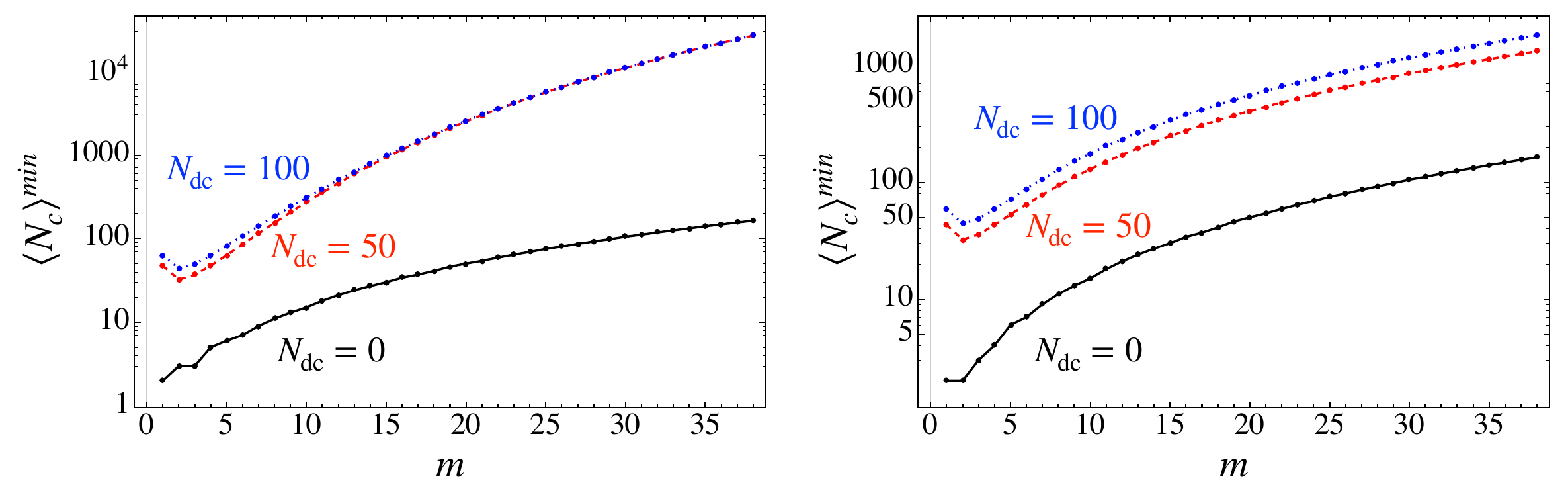}
\caption{Black line and dots: minimum value $\langle N_{c} \rangle^{min}$ such that Eq.~\eqref{eq:NthH} (left plot) and Eq.~\eqref{eq:NthV} (right plot) are satisfied as a function of $m$ in the absence of dark counts ($N_{\Dk} = 0$). Red line and dots ($N_{\Dk} = 50$), blue line and dot ($N_{\Dk}=100$): minimum vale $\langle N_{c} \rangle^{min}$ such that Eq.~\eqref{eq:NcthpoissH} (left plot) and Eq.~\eqref{eq:NcthpoissV} (right) plot are satisfied. Notice the different scaling for the $y$-axis.}
\label{fig:minNc}
\end{figure}

In an ideal scenario, namely, without fluctuations in the sent number of photons, it is clear that the two approaches are complementary and yield to the same conclusion, since the single detections in $H$ and $V$ are perfectly correlated. Moreover, given that only the words $a^{k} \in L_{m}$ satisfy the condition $f_{k}>f_{\rm th}$, with this strategy we have a zero error probability, provided that $\langle N_{\rm c} \rangle$ is large enough such that the integer part of $N_{\rm th}^{H}$ ($N_{\rm th}^{V}$) is strictly positive (negative) than $N^{H}_{\rm c}(k_1)$ ($N^{V}_{\rm c}(k_1)$), i.e. we have the conditions
\begin{subequations}
\begin{align}
\label{eq:NthH}
\floor*{N_{\rm th}^{H}} & = \floor*{\frac{\langle N_{\rm c} \rangle(1+\cos^{2}\theta_{m})}{2}} > \floor*{\langle N_{\rm c}\rangle \cos^{2}\theta_{m}}, \\
\label{eq:NthV}
\floor*{N_{\rm th}^{V}} & = \floor*{\frac{\langle N_{\rm c}\rangle \sin^{2}\theta_{m}}{2}} < \floor*{\langle N_{\rm c}\rangle \sin^{2}\theta_{m}}.
\end{align}
\end{subequations}
In Fig. \ref{fig:minNc} (black lines and dots) we report the minimum values of $\langle N_{c}\rangle$ such that the last two inequalities hold.
\par

In a realistic scenario, the photo-detection is influenced by two distinct noisy effects that can affect the error probability. The first is that the number of detected photons follows a Poisson distribution \cite{loudon2000quantum}, that is we have 
\begin{equation}
\textup{Poi}(n;\mu) = \frac{\mu^{n}e^{-\mu}}{n!}
\end{equation}
that is the probability of detect $n$ photons depends on the average number of detected photons $\mu$. How this affects the error probability has been thoroughly addressed both theoretically and experimentally in \cite{PhysRevResearch.2.013089}.

The second effect that we should consider in order to apply our enhanced strategy is due to the {\it dark counts}, namely, the random counts registered by the detector without any incident light on it. Being still related to the detection process, also the dark counts follow a Poissonian distribution, whose mean $\langle N_{\Dk}\rangle$ depend on the particular detector one choose to use. In a typical quantum optics experiments, the dark-count rate ranges from tens to hundreds of photons per second, but this number can be drastically reduced by using coincidence counting techniques \cite{bachor}, up to making this effect negligible. For instance, in the implementation in \cite{PhysRevResearch.2.013089} the dark counts where only 0.001\% of the effective coincidence counts. Since the dark counts occurs randomly, we cannot distinguish between a dark count and signal one. Therefore, the probability of detecting $N$ photon in the $H$ or $V$ photodetector for a word $a^{k}$ is finally given by
\begin{align}
P^{x}_{k}(N) & = \sum_{n=0}^{+\infty}\sum_{m=0}^{+\infty}\textup{Poi}\left(n;\eta^{x}\right)\textup{Poi}\left(m;\langle N_{\Dk}\rangle\right) \delta_{n+m,N} \\[1ex]
& = \textup{Poi}(N; \mu_{k}^{x})
\label{eq:PoissHVDc}
\end{align}
where $ \eta^{H}= \langle N_{\rm c} \rangle \cos^{2}(k \theta_m)$ and $\eta^{V} = \langle N_{\rm c} \rangle \sin^{2}(k \theta_m)$, while we have defined the {\it overall} mean number of detected photons as 
\begin{subequations}\label{aveHV}
\begin{align}
\mu_{k}^{H} & = \langle N_{\rm c}\rangle \cos^{2}\left(k \theta_m\right) + \langle N_{\Dk}\rangle,\\
\mu_{k}^{V} & = \langle N_{\rm c}\rangle \sin^{2}\left(k \theta_m\right) + \langle N_{\Dk}\rangle.
\end{align}
\end{subequations}

As we noticed above, the dark count rate is usually very small with respect to the detected count rate of the signal. Therefore, for the $H$ detector which detects the higher number of photons, see Eqs.~\eqref{aveHV}, they are relevant only when $\langle N_{\rm c}\rangle \sim \langle N_{\Dk}\rangle$. On the contrary, for the $V$ detector, detecting the lower number of photons, their role is fundamental in determining the performance of the photonic automaton, since $\mu_{m}^{V}= \mu_{\Dk} = \langle N_{\Dk}\rangle$. This is the main difference between the two strategies: in the first, we need to distinguish between two finite mean numbers of photon $\mu^{H}_{m} = \langle N_{\rm c}\rangle + \langle N_{\Dk}\rangle$ and $\mu^{H}_{k_1}$, while in the second case we need to distinguish between the noise due to dark counts, being $\mu_{m}^{V} = \langle N_{\Dk}\rangle$, and $\mu^{V}_{k_1}$. However, to assess the performance of second strategy with respect the first one, we need to evaluate the probability of errors in the two cases.
\par
\begin{figure}
\centering
\includegraphics[width=0.7\textwidth]{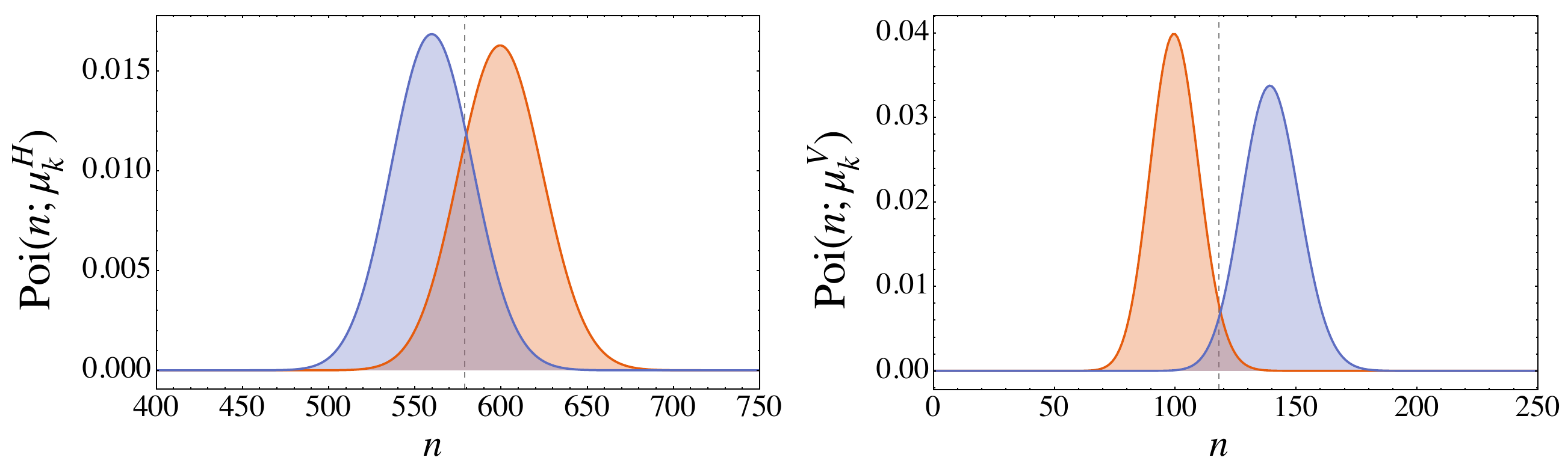}
\caption{Probability density function of the Poissonian distribution in Eq. \eqref{eq:PoissHVDc} for the $H$ detector (left plot) and the $V$ detector (right plot) for $\langle N_{c} \rangle = 500$, $N_{\Dk}=100$ and $m=11$. The probability of error in Eq. \eqref{eq:peV} and \eqref{eq:peH} are respectively $p_{e}^{V}=0.034$ ($V$ detector) and $p_{e}^{H} = 0.205$ ($H$ detector). The grey dashed line is the threshold values in Eq. \eqref{eq:thresholdNthx}. The values of the involved parameters have been chosen to better highlight the investigated effect.}
\label{fig:poisson}
\end{figure}
Let us first find the threshold values in the two different strategy. We need to find the intersection between two Poissonian distributions for a word belong to $L_{m}$ and a word $a^{k_1}$ with highest probability of being erroneously being accepted, as show in Fig.~\ref{fig:poisson}. By imposing  
$\textup{Poi}(N_{\rm th}^{x};\mu^{x}_{1}) = \textup{Poi}(N_{\rm th}^{x};\mu^{x}_{m})$,
where $x=H,V$, we find an exact solution for $N_{\rm th}^{x}$ given by (see the vertical dashed line in Fig.~\ref{fig:poisson})
\begin{equation}
N_{\rm th}^{x} = \frac{\mu^{x}_{m}-\mu^{x}_{k_{1}}}{\ln{\mu^{x}_{m}}-\ln{\mu^{x}_{k_{1}}}}.
\label{eq:thresholdNthx}
\end{equation}
To highlight the dark counts effects, we introduce the ratio $\eta = \langle N_{\Dk} \rangle / \langle N_{\rm c} \rangle$, and we have
\begin{align}
N_{\rm th}^{H} &=
\frac{\langle N_{\rm c}\rangle \sin^{2}\theta_m}{
\ln\left(1+\eta\right)-
\ln\left(\cos^{2}\theta_m + \eta\right)
}, \\[1ex]
N_{\rm th}^{V} &= \frac{\langle N_{\rm c}\rangle \sin^{2}\theta_m}{
\ln\left(\sin^{2}\theta_m + \eta\right)-
\ln\left(\eta\right)
}.
\end{align}
\par
In our framework, the accepting problem is introduced as binary discrimination between the correct word and the word with the highest probability of error. However, in the photonic realization of the automata \cite{PhysRevResearch.2.013089}, when the number of input photons is small and $m$ is large, also word with larger $k (\mbox{mod}~m)$ may contribute to the error. For this reason, like in the ideal case, we establish the minimum number of input photon $\langle N_{\rm c}\rangle^{min}$ which are necessary to faithfully consider the problem as binary discrimination.

To have faithfully binary discrimination the fluctuations due to the word with the second-largest probability of error, i.e. a word $a^{k_{2}}$ with $ k_{2} (\mbox{mod}~m) = 2$, must be much larger than the fluctuations due to the correct word, where here for ``large'' we mean at least two standard deviations. In this way, the discrimination can be considered only between the words $a^{m}$ and $a^{k_{1}}$. In the case of a Poissonian random variable, the standard deviation is the square root of the mean for Poissonian random variables. Hence, we have the two conditions respectively for the $H$ and $V$ detector 
\begin{subequations}
\begin{align}
 \label{eq:NcthpoissH}
 \mu_{k_{2}}^{H} + 2\sqrt{\mu_{k_{2}}^{H}} < \mu_{m}^{H} - 2\sqrt{\mu_{m}^{H}},\\[1ex]
 \label{eq:NcthpoissV}
\mu_{Dc} + 2 \sqrt{\mu_{Dc}} < \mu_{k_{2}}^{V} - 2\sqrt{\mu_{k_{2}}^{V}}.
 \end{align}
 \end{subequations}
In the first one we ask that the fluctuations due to the word with the second largest probability of error, i.e. a word $k_{2}$ with $ k_{2} (\mbox{mod}~m) = 2$ are much larger than the fluctuations due to dark counts. In a similar way we define the threshold for the horizontal detector. These equations can be solved for $\langle N_{\rm c}\rangle$ and set a lower bounds for it such that the probability of error can be evaluated in term of a binary discrimination problem, as shown in Fig.~\ref{fig:minNc} (red and blue lines and points) .
\par
Now we can now evaluate the probability of error for the two strategies. Indeed, this is equal to 
\begin{equation}
p_{e}^{x} = p(a^{k_1})p^{x}(a^{k_1}\to a^{m}) + p(a^{m}) p^{x}(a^{m} \to a^{k_{1}}),
\end{equation}
where we have denoted $p^{x}(a ^{i} \to a^{j})$ as the probability of detecting the word $a^{i}$ as $a^{j}$ by the detector $x = H,V$.
Since we have no a priori knowledge on the input word we set the prior probabilities $p(a^{k_1}) = p(a^{m}) = 1/2$, and for the $V$ detector we obtain
\begin{align}
\label{eq:peV}
P_{e}^{V} & = \frac{1}{2}\left[
\sum_{n=0}^{\floor*{N_{\rm th}^{V}}} \frac{(\mu_{k_{1}}^{V})^{n}e^{-\mu_{k_{1}}^{V}}}{n!} +
\sum_{n= \floor*{N_{\rm th}^{V}}+1}^{+\infty} \frac{\mu^{n}_{\Dk}e^{-\mu_{\Dk}}}{n!} \right] \\[1ex]
& = \frac12 \left[ 1- \frac{\Gamma(\floor*{N_{\rm th}^{V}}+1,\mu_{\Dk})-\Gamma(\floor*{N_{\rm th}^{V}}+1, \mu_{k_{1}}^{V})}{\floor*{N_{\rm th}^{V}}!} \right] \\[1ex]
& = \frac12\left[ 1 - \int_{N_{\Dk}}^{N_{\rm c}\sin^{2}\theta_m + N_{\Dk}}  \frac{e^{-t}t^{\floor*{N_{\rm th}^{V}}}}{\floor*{N_{\rm th}^{V}}!} dt \right] ,
\end{align}
where $\Gamma(a,x)$ is the incomplete Gamma function
\begin{equation}
\Gamma(a,x) = \int_{x}^{+\infty} e^{-t}t^{a-1}dt.
\end{equation}
Analogously, we may evaluate the probability of error for the detection of a horizontally polarized photon, i.e.
\begin{align}
P_{e}^{H} & = \frac{1}{2} \left[ \sum_{n=0}^{\floor*{N_{\rm th}^{H}}} \frac{(\mu_{m}^{H})^{n}e^{-\mu_{m}^{H}}}{n!} + \sum_{n= \floor*{N_{\rm th}^{H}}+1}^{+\infty} \frac{(\mu_{k_{1}}^{H})^{n}e^{-\mu_{k_{1}}^{H}}}{n!} \right].
\label{eq:peH}
\end{align}
\par
Eventually, we can introduce a third strategy that combines the two described so far: for each beam of photon, we propose to measure both the $H$ and $V$ polarization and to combine the results so obtained. From a theoretical point of view, this is equivalent to the automata presented before, since in the ideal case the two detectors perfectly agree, i.e. one sees the photon and the other one does not see it. However, in the non-ideal case, noisy fluctuations affect photodetection. Since the fluctuations in the $H$ detector are independent from the one in the $V$ detector, the probability of erroneously accepting it by looking at both $H$ and $V$ is given as
\begin{equation}
p_{e}^{J} = p(a^{k_{1}}) p^{H}(a^{k_{1}}\to a^{m})p^{V}(a^{k_{1}}\to a^{m}) + p(a^{m}) p^{H}(a^{m}\to a^{k_{1}}) p^{V}(a^{m}\to a^{k_{1}}) 
\end{equation}
where $J$ here stands for \emph{joint}.

\section{Numerical results and simulations}\label{s:numerical}
\begin{figure}
\centering
\includegraphics[width=0.7\textwidth]{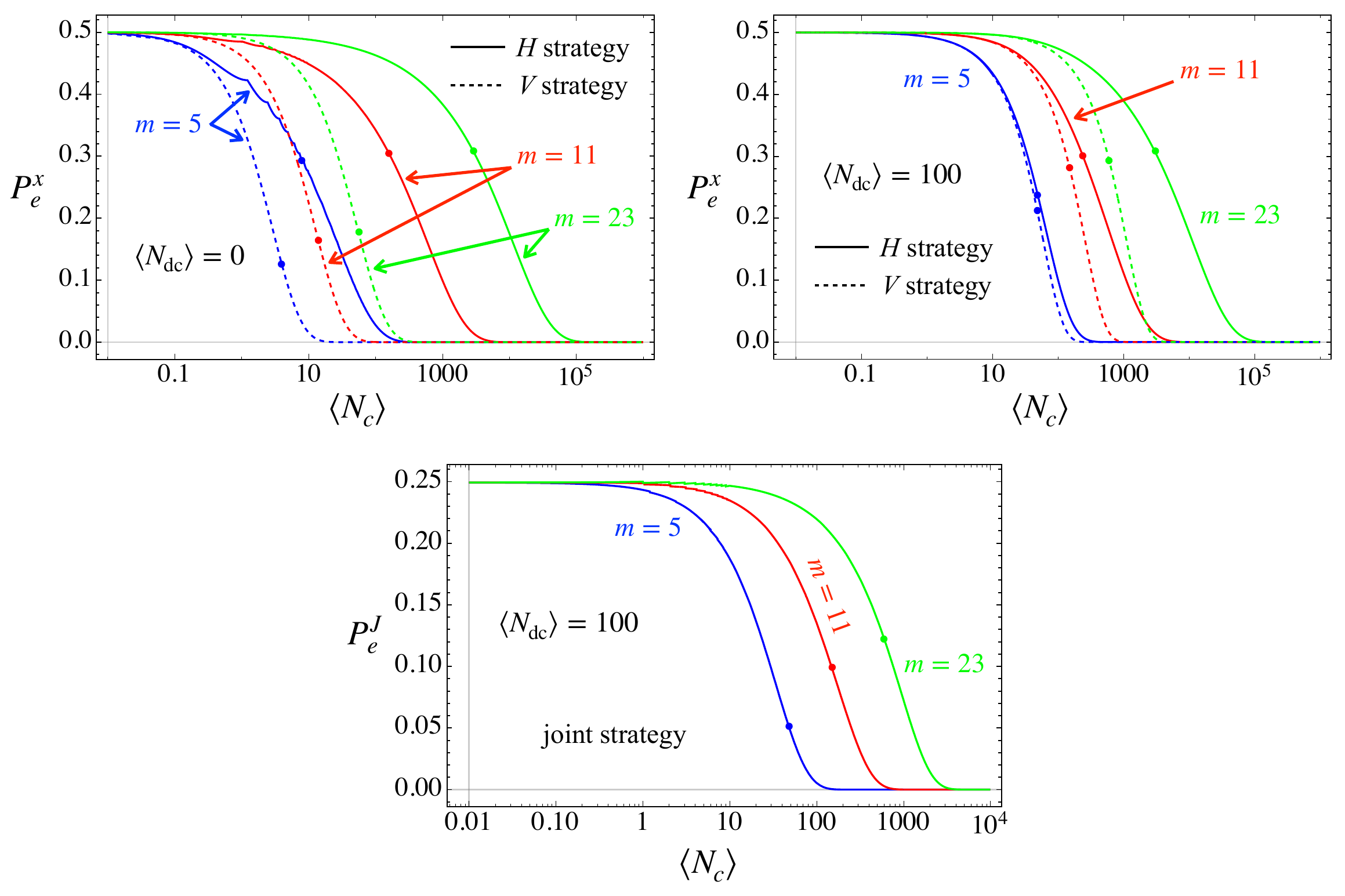}
\caption{Probability of error for the different strategies as functions of the average number of input photon $\langle N_{\rm c}\rangle$ in a semi-log plot. Red line: $m=5$; blue line: $m=11$; green line $m=23$. Top panels: $H$ (solid lines) and $V$ (dashed lines) strategies in the absence of dark counts (left) and for $\langle N_{\Dk}\rangle = 100$ (right). Bottom panel: joint strategy in the case $\langle N_{\Dk}\rangle = 100$. The dots on the lines refer to the threshold values evaluated according to \eqref{eq:NcthpoissH} and \eqref{eq:NcthpoissV}. See the text for details.}
\label{fig:perror1}
\end{figure}

The comparison of the three strategies is reported in Fig. \ref{fig:perror1}. We see that the $V$ strategy outperforms the $H$ strategy for all the possible values of input photon, reaching almost a negligible error for approximately an order of magnitude less than the $H$ strategy. The joint strategy realizes a further enhancement, even though the $p_{e}^{J}$ approaches $0$ with the same order of magnitude of $\langle N_{\rm c}\rangle $ as $p_{e}^{V}$. We have also reported the solution for the inequalities  \eqref{eq:NcthpoissH} and  \eqref{eq:NcthpoissV} as a point along the corresponding line: for smaller value, the probabilities of error are not reliable since the contribution of the words with larger $k (\mbox{mod}~m)$ is not negligible. In addition, increasing the average number of dark counts slightly increases the probability of error for all the strategies considered, even though no significant effects are detected for the considered range of values of $\langle N_{\Dk} \rangle$.
\begin{figure}
\centering
\includegraphics[width=0.7\textwidth]{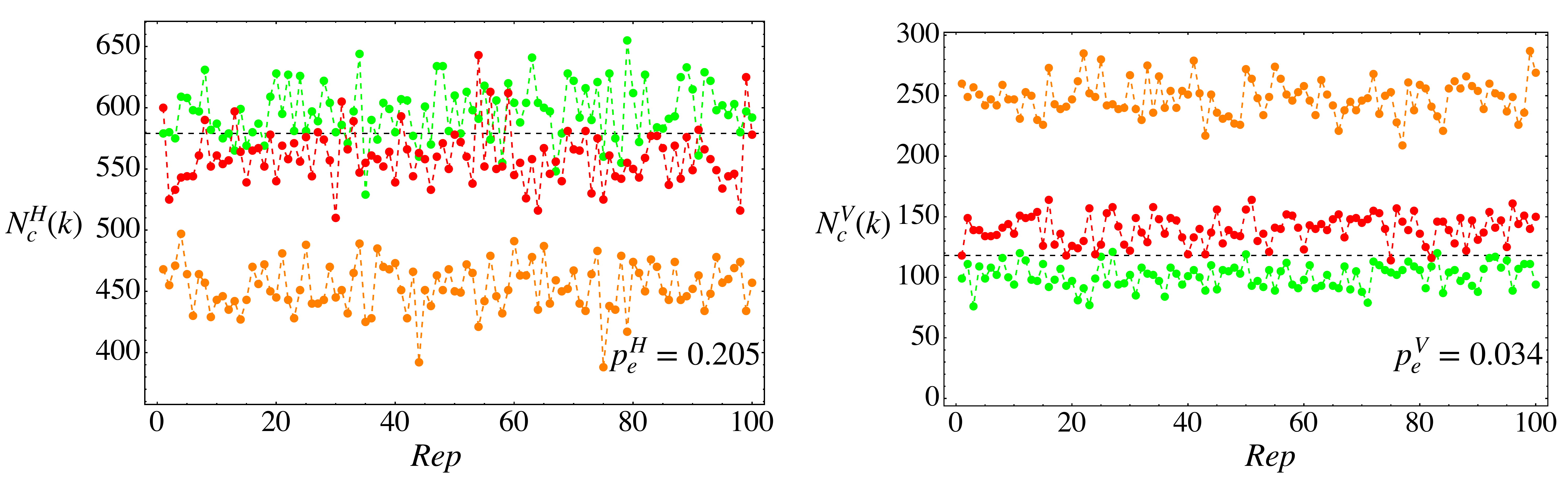}
\caption{Simulation of $N_{\rm c}^{x}(k)$ for the horizontal (left) and vertical (right) automata as a function of the experimental run number ($Rep$). Green dot: $k=m=11$, i.e. $k (\mbox{mod}~m) = 0$; red dot: $k= 12$, i.e. $k (\mbox{mod}~m) = 1$;  orange dot: $k= 13$, i.e. $k (\mbox{mod}~m) = 2$; black dashed line: $N_{\rm th}^{x}$. We considered $\langle N_{\Dk} \rangle = 100$ and $\langle N_{\rm c}\rangle = 500$ (the same parameters of Fig. \ref{fig:poisson}). The probabilities of error given in Eq. \eqref{eq:peV} and Eq. \eqref{eq:peH} are respectively $p_{e}^{V}=0.034$ and $p_{e}^{H} = 0.205$. The minimum number of input $\langle N_{\rm c}\rangle$ for the $H$ detector, solution of \eqref{eq:NcthpoissH}, is $\langle N_{\rm c}^{H} \rangle^{min}=238$, while for $V$, solution of \eqref{eq:NcthpoissV}, is $\langle N_{\rm c}^{V} \rangle^{min}=151$.}
\label{fig:repHV}
\end{figure}
\par
In Fig.~\ref{fig:repHV} we show the number of counts at the $H$ and $V$ detectors from a simulated experiment. We can see a significant reduction of the fluctuations in the $V$ detector, which is also marked by the significant difference in the probability of error $p_{e}^{H}$ and $p_{e}^{V}$. The main reason is that the counts in the $V$ detector are affected only by the randomness due to the dark counts (if present), while in the $H$ detector the expected number of photons contributes to the randomness of the outcomes as well. We have also reported the results for words of length $k (\mbox{mod}~m) = 2$, which are are significantly separated from $N_{\rm c}^{x}(m)$ and $N_{\rm c}^{x}(k_{1})$ since the value of $\langle N_{\rm c}\rangle$ considered is much larger than the threshold given in \eqref{eq:NcthpoissH} and \eqref{eq:NcthpoissV}.

\section{Conclusions}\label{s:concl}

In this work we have presented an enhanced photonic implementation of 1qfa for the recognition of unary language that significantly improves the performance obtained by the one originally proposed in \cite{PhysRevResearch.2.013089}. The protocol uses the polarization degree of freedom of single photons, and exploits the possibility of detecting not only the horizontal polarization, as in \cite{PhysRevResearch.2.013089}, but also the vertical one. The resulting scheme largely outperforms the original automaton for smaller values of the mean number of sent photon $\langle N_{c}\rangle$. In addition, we have extended the results previously found with a detailed analysis of the conditions for which such 1qfa can work with high reliability. We have evaluated the minimum number of photons that must be sent in order to solve faithfully the inherent binary discrimination problem. As one would expect, the minimum $\langle N_{c} \rangle$ is smaller for the automaton that relies on the new strategy based on the $V$ detector.

In our analysis we have discussed the presence of dark counts in the detection of both strategies, we have evaluated their effects both on the probability of error and on the minimum $\langle N_{c}\rangle$. Eventually, we have also examined a joint strategy in which we combine both the $H$ and the $V$ detection, which can indeed be used at no additional cost. We have therefore proved that when the number of sent photon is constrained to small values, the $V$ detection version of the 1qfa should be preferred.

Our results pave the way to the effective implementation of 1qfa using quantum optical platform, thus opening the possibility of processing strings of input symbols using feasible devices and, in turn, to introduce quantum languages and compare the complexity of classes of languages in classical and quantum cases. More generally, since the assessment of the actual power of quantum computers is one of the most significant challenges of quantum technology, implementing quantum automata provides a relevant arena
to better understand the computing capabilities offered by quantum devices.

\acknowledgments{M.~G.~A.~Paris is member of GNFM-INdAM. C.~Mereghetti and B.~Palano are members of GNCS-INdAM. We thank V.~Vento for useful discussion.}

\end{document}